\documentclass[aps,pra,showpacs,preprint,superscriptaddress,amsmath,amssymb, floatfix]{revtex4-1}
\usepackage{graphicx}
\usepackage{epstopdf}
\usepackage{subfigure}
\usepackage{dcolumn}
\usepackage{bm}
\usepackage{extarrows} 
\usepackage{MnSymbol}
\usepackage{hyperref}
\hypersetup{colorlinks=true,       
   linkcolor=blue,          
   citecolor=blue,        
   urlcolor=blue }	

\def\be{\begin{equation}}      
\def\ee{\end{equation}}
\def\bea{\begin{eqnarray}}      
\def\eea{\end{eqnarray}}

\def\beu{\begin{equation*}}   
\def\eeu{\end{equation*}}
\providecommand{\bv}[1]{\boldsymbol{#1}}
\providecommand{\mr}[1]{\mathrm{#1}}
\providecommand{\abs}[1]{\left\lvert#1\right\rvert}   
\providecommand{\ket}[1]{\left|#1\right\rangle}

\providecommand{\bra}[1]{\left\langle#1\right|}

\providecommand{\ensemavg}[1] {\left\llangle #1\right \rrangle }

\providecommand{\pp}[2]{\frac{\partial{#1}}{\partial{#2}}}

\begin{document}
\title{Optomechanically-induced chiral transport of phonons in one dimension}


\author{Xunnong Xu }
\affiliation{Joint Quantum Institute, University of Maryland/National Institute of Standards and Technology, College Park, Maryland 20742, USA}
\author{Jacob M. Taylor}
\affiliation{Joint Quantum Institute, University of Maryland/National Institute of Standards and Technology, College Park, Maryland 20742, USA}
\affiliation{Joint Center for Quantum Information and Computer Science, University of Maryland, College Park, Maryland 20742, USA}
\affiliation{Research Center for Advanced Science and Technology (RCAST), The University of Tokyo, Meguro-ku, Tokyo, 153-8904, Japan}



\date{\today}
\begin{abstract}
Non-reciprocal devices, with one-way transport properties, form a key component for isolating and controlling light in photonic systems. Optomechanical systems have emerged as a potential platform for optical non-reciprocity, due to ability of a pump laser to break time and parity symmetry in the system. Here we consider how the non-reciprocal behavior of light can also impact the transport of sound in optomechanical devices. We focus on the case of a quasi one dimensional optical ring resonator with many mechanical modes coupled to light via the acousto-optic effect. The addition of disorder leads to  finite diffusion  for phonon transport in the material, largely due to elastic backscattering between clockwise and counter-clockwise phonons. We show that a laser pump field, along with the assumption of high quality-factor, sideband-resolved optical resonances, suppresses the effects of disorder and leads to the emergence of chiral diffusion, with direction-dependent diffusion emerging in a bandwidth similar to the phase-matching bandwidth for Brillouin scattering. A simple diagrammatic theory connects the observation of reduced mechanical linewidths directly to the associated phonon diffusion properties, and helps explain recent experimental results.
\end{abstract}


\maketitle

\section{Introduction} 
Optomechanics, in the form of directional light-matter coupling via Brillouin scattering as a result of momentum conservation, leads to non-reciprocal behavior for light propagation \cite{Manipatruni2009, Hafezi2012, Dong2015, Kim2015, Shen2016} when a pump laser breaks time and parity symmetry. These studies focused on new ways of building on-chip optical isolation device \cite{Yu2009, Bi2011, Huang2016}, with applications in optical quantum computation \cite{Kok2007, OBrien2007} and quantum simulations \cite{Hafezi2013}. Similarly, we posited that the directional optomechanical interaction generically leads to non-reciprocal phonon behavior \cite{Xu2016chirality}. This non-reciprocal behavior may show up in a variety of settings. For example, in the presence of phonon scattering via disorder, impurities, or phonon-phonon interactions,  phonons traveling in both directions undergo reciprocal, diffusive transport. In particular, disorder does not conserve parity. However, in a recent paper \cite{Kim2016nature}, the parity-conserving,  chiral transport of phonons was observed in a whispering-gallery type microsphere resonator in the optically non-reciprocal regime. This is due to acoustic-optic coupling with a strong, directional pump, even in the presence of phonon scattering via impurities. That experiment showed that the linewidth of clockwise and counter-clockwise phonons can be modified in opposite ways at the same time.

A minimal theory for explaining the optically-induced chiral behavior of phonons observed in \cite{Kim2016nature} is possible with just a few assumptions \cite{Xu2016chirality}. However, understanding how optically-induced chirality impacts phonon transport in the continuum regime requires a more detailed understanding of the modification of the phonon propagation, which we provide here. We assume that there are two counter-propagating optical modes (control and probe) and a continuum of phonon modes (including both forward and backward propagating modes) in the system, so Brillouin scattering can be stimulated \cite{Smith1972, Ippen1972}. We assume narrow optical resonances (sideband-resolved regime) to prevent anti-stokes phonon generation processes.The directional coupling to the optical fields provided by the phase-matching condition of Brillouin scattering optomechanically cools phonons in one direction \cite{Mancini1998, Marquardt2007}, while phonons in the other direction do not experience this optical modification directly because they are not phase matched. When considering a series of random phonon scattering processes which cause phonon diffusion \cite{Datta1997}, we show that the asymmetry in the free phonon propagation due to optomechanical coupling eventually leads to chiral behavior for phonon transport.

Our theory is an expansion and extension of our previous work \cite{Xu2016chirality} in collaboration with experimental efforts \cite{Kim2016nature}. In contrast to those works, here we directly consider the many-mode behavior of the system and use a simple diagrammatic perturbation theory to connect experimentally measurable quantities in the optics to the phonon diffusion. In Sec II, we consider a model where multiple phonon modes are coupled to a common optical mode with different weights, due to phase matching. We also consider phonon scattering via impurities, with the introduction of a random scattering potential. Subsequently, in Sec III, we use the Heisenberg-Lagevin equations to find the linear response of phonons. In particular,  we consider the case where only a few phonon modes are strongly coupled to the optical mode, and we use numerical simulation of the scattering potential to find the linear response function. Finally, in Sec IV, we use diagrammatic perturbation expansion in disorder potential to calculate the phonon linear response of a multi-mode coupled system to find the phonon linewidth, and connect the experimentally observable phonon mode quality (Q) factor to the phonon diffusion rate.


\section{Multi-mode theory of the elastic-optic interaction in the sideband-resolved regime} 
\begin{figure}[!htp]
\begin{center}
\includegraphics[width=0.9\columnwidth]{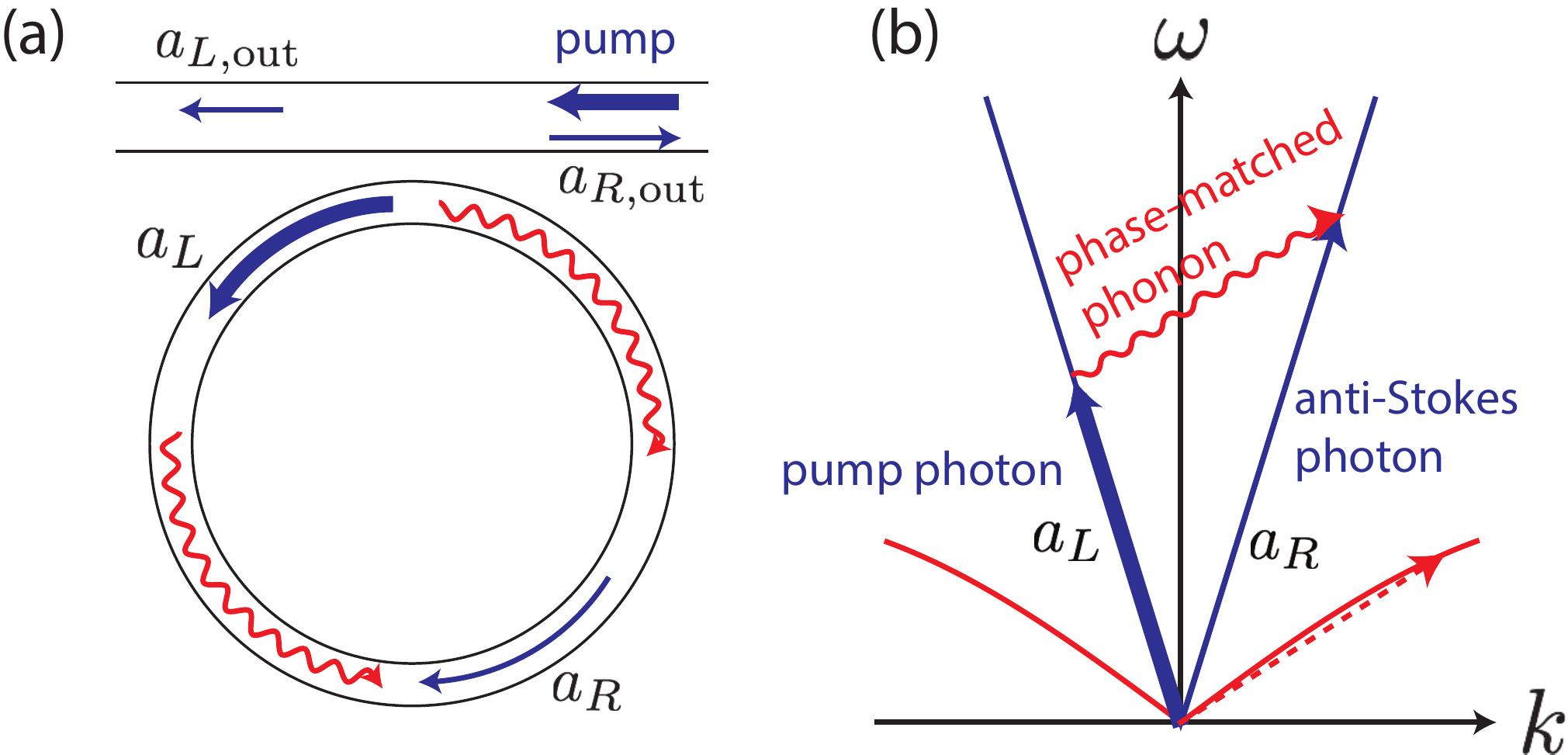}
\caption{(color online). (a) Schematic of the ring waveguide resonator that supports two counter-propagating optical modes $a_{\mathrm{L}}$, $a_\mathrm{R}$ and continuum of phonon modes $b_\mathrm{\bf{q}}, b_\mathrm{-\bf{q}}$. (b) The dispersion relation curves for photon (blue line) and acoustic phonon (red curves). An pump photon absorbs an incoming phonon and then produce a counter-propagating anti-Stokes photon with a higher energy. Both momentum and energy are conserved in this anti-Stokes Brillouin scattering process.}
\label{fig:fig1}
\end{center}
\end{figure}

We consider a continuum theory for a quasi one dimensional  waveguide resonator that supports photons and phonons at the same time.  We focus on the one dimensional case for its simplicity and transparency, though we note that our motivation and prior, simpler theoretical work examined the three dimensional case \cite{Kim2016nature, Xu2016chirality}. The phonons are subject to periodic boundary conditions (PBC) and the photons are send into the waveguide through evanescent coupling to an  optical waveguide.  The control field , which sets the directional propagation via phase matching and Brillouin scattering, is red detuned from a counter-propagating probe optical mode with the detuning $\Delta=\omega_c - \omega_p <0$, and we will assume that optical modes are narrow enough in frequency to be sideband-resolved ($\kappa < \Delta$). Brillouin scattering is possible by absorbing a phonon and a control photon and subsequently creating an anti-Stokes probe photon, as shown in Fig.~\ref{fig:fig1}.  

The  phonon modes in the waveguide are the second quantized form of the normal modes of the displacement field of the material. In the quasi-1D limit, when the dimension of the cross section of the waveguide is small compared to the radius of the ring $r_0$, the displacement field in the longitudinal direction is described in polar coordinates by  $\bv{u} = u(\theta,t)$. The Lagrangian of the system reads:
\bea
L = \int_0^{2\pi} \mathcal{L_{\theta}} \mr{d}\theta, 
\eea
where the Lagrangian density is 
\be
\mathcal{L_{\theta}} =  \frac{1}{2} \left[ \rho_{\theta} (\partial_t u)^2 - \frac{Y}{r_0} (\partial_\theta u)^2 \right], 
\ee
$\rho_\theta$ is the density per unit angle, and $Y$ is the Young's modulus. The conjugate momentum density is defined as 
\be
\pi = \pp{\mathcal{L_\theta}}{(\partial_t u)} = \rho_\theta \partial_t u,
\ee
so the Hamiltonian density is 
\be
\mathcal{H} = \pi \partial_t u - \mathcal{L_\theta} = \frac{1}{2} \left[ \frac{\pi^2}{\rho_{\theta}} + \frac{Y}{r_0} (\partial_\theta u)^2 \right]. 
\ee

We now go to momentum space and expand $u,\pi$ in normal modes: 
\begin{subequations}
\bea
u(\theta, t) &=& \sum_q u_q e^{iqr_0\theta},\\
\pi (\theta, t) &=& \sum_q \pi_q e^{iqr_0\theta}. 
\eea
\end{subequations}
with the commutation relations $[u_q, \pi_{q^{\prime}} ] = i \hbar \delta_{q, q^{\prime}} $. 
The Hamiltonian density is then integrated
\bea
H = \int_0^{2\pi} \mathcal{H} \mathrm{d} \theta =2\pi \sum_q \left( \frac{ \pi_q \pi_{-q} }{2\rho_\theta}  + \frac{1}{2}\rho_\theta \Omega_q^2 u_q u_{-q} \right)
\eea
where $\Omega_q = q v$ is the frequency of mode $q$ and $v = \sqrt{Y r_0/\rho_\theta}$ is the (angular) speed of sound. Since the position-space $\pi$ and $u$ are both hermitian operators, we have $\pi_q^{\dag} = \pi_{-q}$ and $u_q^{\dag} = u_{-q}$.  We  define the following annihilation operator 
\begin{subequations}
\bea
b_q &=& \sqrt{\frac{\rho_\theta\Omega_q}{2\hbar}} \left(u_q + i \frac{1}{\rho_\theta\Omega_q} \pi_q\right).
\eea
\end{subequations}
where $x^{q}_{\mathrm{zpf}} = \sqrt{\rho_\theta \Omega_q/2 \hbar}$ is the size of the phonon ground state in $\theta$. 
The Hamiltonian can be described by a set of decoupled quantum harmonic oscillators, namely phonons modes 
\be
H_{\mr{phonon}} =  \sum_q \Omega_q b_q^{\dag} b_q.  \quad\quad (\hbar=1)
\ee
We note that the phonon group velocity $v$ is assumed constant, because we only consider the long wavelength phonons in the accoustic branch, as show in Fig.~\ref{fig:fig1}.

The electric field of the optical mode in the resonator can also be expanded  as \cite{Scully1997}
\be
E(t, \theta) = \sum_k \sqrt{\frac{\hbar\omega_k}{2\epsilon V}} a_k e^{ik r_0\theta} + \mathrm{h.c.}, 
\ee
where $\epsilon$ is the susceptibility of the medium, $V=2\pi r_0 A$ is the volume of the waveguide, $a_k$ is the annihilation operator for mode $k$ in the Heisenberg picture, and $\omega_k/2\pi$ is its frequency. The Hamiltonian  of the optical field in second quantized form is
\be
H_\mathrm{photon} = \sum_k \omega_k a_k^{\dag} a_k,
\ee
which represents a sum of independent quantum harmonic oscillators. Again, for simplicity we consider a single optical transverse mode. Using periodic boundary conditions, we find that $\omega_k=\abs{n_k} c/r_0 $, with $n_k \in \mathbb{Z}$ the mode number. The dispersion relation can be of a more general form, but the most important part is the phase in Eq. (9), which is related to phase-matching, as we will discuss in more detail later. 

The interaction between the mechanical motion and electromagnetic field comes from acousto-optic effect: a change in the susceptibility of the material because of the strain from the displacement. This further leads to the change in the energy of the electromagnetic field inside the material:
\be
H_\mathrm{int} \approx \int \frac{1}{2} \pp{\epsilon}{s} s \abs{\mathcal{E}}^2 r_0 A \mr{d}\theta. 
\ee
The strain field $s$ is related to the displacement field $u$ by the definition $s \equiv \partial_\theta u/r_0$, and we take $\partial \epsilon/\partial s$ to be a constant determined by material properties. In explicit form, the interaction is  
\begin{widetext}
\bea
H_\mathrm{int} &=&  \frac{1}{2} A \pp{\epsilon}{s} \int \sum_q iq r_0 u_q e^{iqr_0\theta} \sum_{k,k^{\prime}}\frac{\hbar\sqrt{\omega_k \omega_{k^{\prime}}} }{2\epsilon V}\left[a_k a_{k^{\prime}} e^{i(k + k^{\prime})r_0 \theta } +  a_k a_{k^{\prime}}^{\dag} e^{i(k - k^{\prime})r_0 \theta } + \mathrm{h.c.} \right] \mr{d}\theta  \nonumber \\
&=& \frac{i \hbar}{8\pi \epsilon} \pp{\epsilon}{s} \int \sum_{q, k, k^{\prime}}   q  \sqrt{\omega_k \omega_{k^{\prime}}}  \sqrt{\frac{\hbar}{2\rho_\theta\Omega_q}} \left(b_q + b_{-q}^{\dag}\right) e^{iqr_0\theta}\left[a_k a_{k^{\prime}} e^{i(k + k^{\prime})r_0 \theta  } +  a_k a_{k^{\prime}}^{\dag} e^{i(k - k^{\prime})r_0 \theta } + \mathrm{h.c.}  \right] \mr{d}\theta  \nonumber \\
&\approx & i\frac{\hbar\omega_k}{8\pi} x_\mr{zpf}^q \frac{\partial\epsilon}{\epsilon_0 \partial s} \sum_{q k k^{\prime}} q(b_q a_k a_{k^{\prime}}^{\dag} + b_{q}^{\dag}a_k^{\dag} a_{k^{\prime}}) f_{k k^{\prime}} (q).
\eea
\end{widetext}
In the last line, the function $f_{k k^{\prime}} (q)$ represents the momentum and energy conservation conditions for forward Brillouin scattering (FBS), corresponding to integrating out $\theta$ and assumption of our effective 1D system. 

In the above calculation, however, we ignored the effects of damping (in the form of non-orthogonal waves with an imaginary part, as arises in quasi-mode theory) and misalignment of the optical and acoustic transverse modes, both of which give $f_{k k^{\prime}} (q)$ a finite width in momentum space.. We will show that inclusion of these effects is self consistent in the final section of the paper -- for now, we leave it as an assumption. Looking only at the damping term, we have  an expected phonon damping from diffusion of $\gamma = D q^2$, where $D$ is the phonon diffusion constant. We neglect the q dependence in $\gamma$ in what follows as we assume the optical system is sideband-resolved, and changes in $q$ away from the phase matching condition are small compared to the overall magnitude of $q$. There is also an imaginary term in the photon wavevector, which is related to the optical loss $\kappa$. Thus in general,  the integration over the spatial coordinate $\theta$ is not a $\delta$-function anymore, but should be replaced by the following function that has a finite width: 
\be
f_{k k^{\prime}} (q) = \int_0^{2\pi} e^{i[ (k -i\frac{\kappa}{2c}) - (k^{\prime} -i\frac{\kappa}{2c}) + (q - i \frac{\gamma}{2v}) ] r_0 \theta} \mathrm{d} \theta = \frac{\gamma/2v + \kappa/c }{(\gamma/2v + \kappa/c) + i(\delta q) }, 
\ee
where $\delta q = q - \Delta k = q - (k^{\prime} -k) $ and we have specialized to the case of two modes: control (with wavevector $k$) and probe (with wavevector $k^{\prime} $).

With this general theory in place, we can now consider the following scenario: we drive the control mode $a_p (\equiv a_k) $ strongly with a red-detuned laser and probe the counter-propagating anti-Stokes mode $a (\equiv a_{k^{\prime}})$  with a weak field. We then linearize the Hamiltonian following the usual optomechanics prescription of replacing $a_p$ with its steady state amplitude $\alpha= \mathcal{E}/(-i\Delta+\kappa/2)$, where the input field amplitude is related to the input power by $\mathcal{E} = \sqrt{P_\mr{in}\kappa/\hbar\omega_k}$. The  Hamiltonian for the interacting system is 
\be
H = -\Delta a^{\dagger}a + \sum_q \Omega_q b_q^{\dagger} b_q + \sum_q i c_{\mathrm{cl}} q\left[f(q) a^{\dagger} b_q - f^*(q) b_q^{\dagger} a\right],
\ee
where and the linearized coupling (angular) speed is given by
\be
c_{\mathrm{cl}} = \frac{\hbar\omega_k}{8\pi} x_{\mr{zpf}}^q \frac{\partial\epsilon}{\epsilon_0\partial s} \alpha.
\ee
The $\alpha$ here is taken to be real, which can be done by a simple gauge transformation without loss of generality. 
We consider the case when the effective detuning of the anti-Stokes mode is close to the frequency of the phonon band: $\Delta = \omega_c - \omega_{p}   \approx -\Omega_q $. 

So far, we have neglected the presence of disorder/defects in the material, which causes phonon scattering and thus mixes phonons with different momenta. We model the defects as random fluctuations of the density of the material $\delta\rho$. This introduces an extra term in the Hamiltonian of the system:
\bea
H_\mathrm{defect} &=& \frac{1}{4\rho_0} \sum_{q q^{\prime}}  \sqrt{\Omega_q \Omega_{q^{\prime}}} (b_q - b_{-q}^{\dag}) (b_{q^{\prime}} - b_{-q^{\prime}}^{\dag}) \int_0^{2\pi} \delta\rho(\theta) e^{i(q+q^{\prime})r_0 \theta} \mathrm{d} \theta  \nonumber \\
&=& -\frac{1}{2\rho_0}  \sum_{q q^{\prime}}  \sqrt{\Omega_q \Omega_{q^{\prime}}}  b_q b_{q^{\prime}}^{\dag} \delta\rho_{q-q^{\prime}} \nonumber \\
&=& - \sum_{q q^{\prime}} g_{qq^{\prime}} b_q^{\dag} b_{q^{\prime}}
\eea
The scattering strength $g_{q q^{\prime}} $ between phonon modes can be calculated numerically by modeling the density fluctuations $\delta\rho(\theta)$ as a quasi-static, random gaussian function, with $\ensemavg{\delta\rho} = 0$ and correlation function $\ensemavg{\delta\rho(\theta) \delta \rho(\theta^{'})} = \bar{U}^2 \delta (\theta-\theta^{'})$, where $\bar{U}$ is the nominal strength of the disorder and the $\ensemavg{\cdot}$ symbols indicate averaging over disorder. We note that from this expression for $g$, we immediately realize that $g_{qq} \propto \delta\rho_{qq} = 0$ for $q\neq 0$, which means that the scattering potential always mixes phonons with different wavevectors. Thus the effect of the disorder will only be to scatter phonons into different outgoing momenta. As we will show, this will lead to diffusive transport for phonons in the absence of optical pumping.

\section{Linear response theory and comparison to prior work} 

Including the damping for sound and light, and the optomechanical interaction in the pump-enhanced frame, we can now describe the linear response of phonons using the Heisenberg-Langevin equations in the input-operator formulation:
\begin{subequations}
\bea
\dot{a} &=& -i(-\Delta-i\kappa/2) a + \sqrt{\kappa}a_{\mr{in}} + \sum_q c_{\mr{cl}} q f(q)b_q, \\
\dot{b}_q &=&  -i(\Omega_q - i\gamma_q/2) b_q + \sqrt{\gamma_q} b_q^{\mr{in}} - c_{\mr{cl}} q f^{*}(q) a + i \sum_{q^{\prime}} g_{q^{\prime} q} b_{q^{\prime}}. 
\eea
\end{subequations}
Here we only consider a single optical mode (that responsible for sideband cooling), and neglect the other probe mode.
We can then move to frequency space and get a matrix equation for the mechanical modes  after  eliminating the optical mode by solution of Eq.~(17a). We get 
\be
D \vec{b}(\omega)= -i \sqrt{\Gamma}  \vec{b}_{\mr{in}}(\omega) + M \vec{b}(\omega) + E \vec{b}(\omega).
\ee
We define $\vec{b} = [b_1, b_2, \cdots, b_q, \cdots]^T$ and the following matrices: 
\begin{subequations}
\bea
\Gamma_{q q^{\prime}}& =& \gamma_q \delta_{q q^{\prime}} , \\
D_{q q^{\prime}}(\omega) &=& (\Omega_q - \omega - i\gamma_q/2) \delta_{q q^{\prime}} , \\
M_{q q^{\prime}}(\omega) &=& \frac{c_{\mr{cl}}^2 q q^{\prime} f(q) f^{*}(q^{\prime})}{(-\Delta-\omega) - i \kappa/2} ,\\
E_{q q^{\prime}} &=&  g_{q q^{\prime}  }. 
\eea
\end{subequations} 
Here $D$ is the bare phonon inverse susceptibility, $M$ represents the effects of the optical mode in both adding damping and mixing mechanical modes, and $E$ is the disorder-induced scattering. 
The optical noise $a_{\mathrm{in}}$, corresponding to a vacuum input field, is usually very small compared to the thermal noise $b_{\mathrm{in}}$, and here we will neglect it. So we find that 
\be
\vec{b}(\omega) = \chi(\omega) (-i\sqrt{\Gamma})  \vec{b}_{\mr{in}}(\omega)= \frac{-i \sqrt{\Gamma}}{D-(M+E)} \vec{b}_{\mr{in}}(\omega), 
\ee
where $\chi(\omega)$ is the susceptibility matrix for the mechanical system.
Since the scattering rate $g_{q q^{\prime}}$ is random, we can find the response matrix by taking the ensemble average of many density fluctuation configurations (again denoted by $\ensemavg{\cdot}$), 
\be
\bar{\chi}(\omega) \equiv \ensemavg{\frac{1}{D-(M+E)} } . 
\ee
The diagonal element of the response matrix $\bar{\chi}(\omega)$ gives the linear response of each mode: 
\be
b_q(\omega) = \bar{\chi}_{qq}(\omega) b_q^{\mr{in}}(\omega). 
\ee

We first consider the case when the mode spacing of phonons is greater than phonon damping rate near the phase-matched condition for sideband cooling: $\Delta q > \gamma/v$. In this regime, only one particular phonon mode interacts resonantly with the optical mode, which means  the main feature of the multi-mode theory is captured in a single-mode minimal theory, as described in Fig.~2(a). We can easily write down the Heisenberg-Langevin equations \cite{Scully1997, Gardiner2004, Walls2008}  for each mode and solve the equations in frequency domain.  From its linear response function, we can extract the approximate damping rates for each mode near resonance $\omega\approx \omega_m$, 
\begin{subequations}
\bea
\gamma_{+}  &\approx & \gamma_{\mathrm{in}} + \gamma_{\mathrm{opt}} + \frac{4g^2}{\gamma_{\mathrm{in}}},  \\
\gamma_{-} & \approx &  \gamma_{\mathrm{in}}   + \frac{4g^2}{\gamma_{\mathrm{in}} + \gamma_{\mathrm{opt}}},
\eea
\end{subequations}
where $+$ is for the forward propagating (CW) phonon and $-$ is for the backward propagating (CCW) phonon.
When $\Delta\approx -\omega_m$, the optomechanical damping rate is $\gamma_{\mathrm{opt}} \approx 4\alpha^2/\kappa$ in the side-band resolved regime, where $\alpha$ is the pump enhanced optomechanical coupling rate. It is clear that when the pump strength $\alpha$ increases, the optomechanical damping rate $\gamma_{\mathrm{opt}}$ also increases, and this leads to broader linewidth for forward propagating phonon \cite{WilsonRae2007, Marquardt2007} and narrower linewidth for backward propagating phonon \cite{Kim2016nature} at the same time.

To compare our multi-mode theory to the simpler theory used in \cite{Kim2016nature, Xu2016chirality}, we choose a few phonon modes and run a numerical simulation to directly calculate the linear response function by taking ensemble average of the random scattering potential. The numerical result is shown in Fig.~2, which is in good agreement with the analytical result for linewidth in Eq.~(23). 
\begin{figure}[h]
\begin{center}
\includegraphics[width=0.75\textwidth]{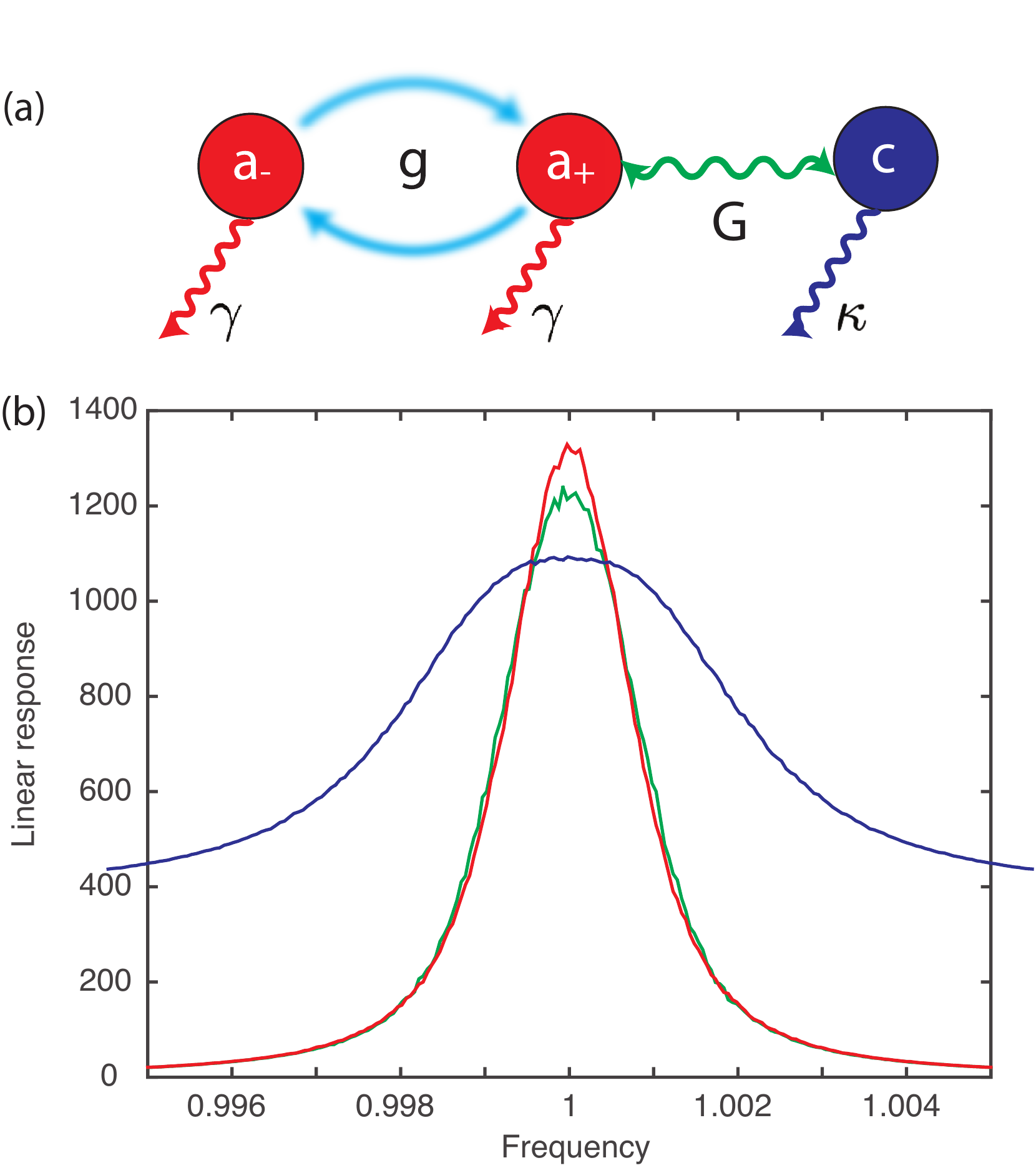}
\caption{Linear response of different phonon modes as a function of rescaled frequency.  Green curve represents the response of phonons without the optomechanical coupling, red (blue) curve represents the response of backward (forward) propagating phonon when the optomechanical coupling is turned on. }
\label{fig:fig2}
\end{center}
\end{figure}

\section{Diagrammatic expansion}
On the other hand, when the phonon mode spacing is smaller than phonon damping rate $\Delta q < \gamma/v$, multiple phonons contribute. We  need to find out how the linewidth of phonon mode depends on the pump power in this multi-mode regime, even as it approaches the continuum limit.  In the following, we setup a formal calculation of linear response function using a simple diagrammatic perturbation expansion. Our goal is still the valuation of the following ensemble-averaged phonon propagator:
\be
\ensemavg{G} = \ensemavg{ \frac{1}{D-(M+E)}} ,
\ee
which would be able to tell us the frequency shift and modified linewidth of each phonon mode. 

\subsection{Expansion of the phonon propagator}
We start by expanding the denominator inside the ensemble average of  Eq.~(24) up to second order in $E$:
\bea
( D - M - E)^{-1} &=& (D-M)^{-1}  + (D-M)^{-1} E (D-M)^{-1} \nonumber \\
&& + (D-M)^{-1} E (D-M)^{-1} E (D-M)^{-1}  + \cdots  
\eea
Because of the assumed stationarity ($\ensemavg{E} = 0$) of the scattering potential, only terms with even numbers of $E$ have a nonzero ensemble average \cite{Datta1997}. To illustrate the interplay between optical cooling and disorder scattering, we define a modified $\tilde{M}$ matrix via 
\bea
\ensemavg{ \left[ D - M - E)\right]^{-1} } & =& (D - \tilde{M}) ^{-1} 
\eea
We first keep only  first order terms in the operator $M$ and second order in $E$, and write
\be
(D-\tilde{M})^{-1} \approx (D-M)^{-1} + (D-M)^{-1} \ensemavg{E (D-M^{-1}) E}  (D-M)^{-1}
\ee
We also can take high order corrections into account, and with a partial summation described in more detail in the next section, we write the result as the following:
\be
\tilde{M} \approx  M + \ensemavg{E \frac{1}{D - M} E}  
\ee 
We see that the modification to the free propagation of phonons has two contributions: the direct optomechanical interaction $M$ and two successive scatterings $E$ with a optomechanically modified phonon propagation $1/(D-M)$ in between, as shown in the diagram below in Fig.~3. 
\begin{figure}[h]
\begin{center}
\includegraphics[width=0.85\textwidth]{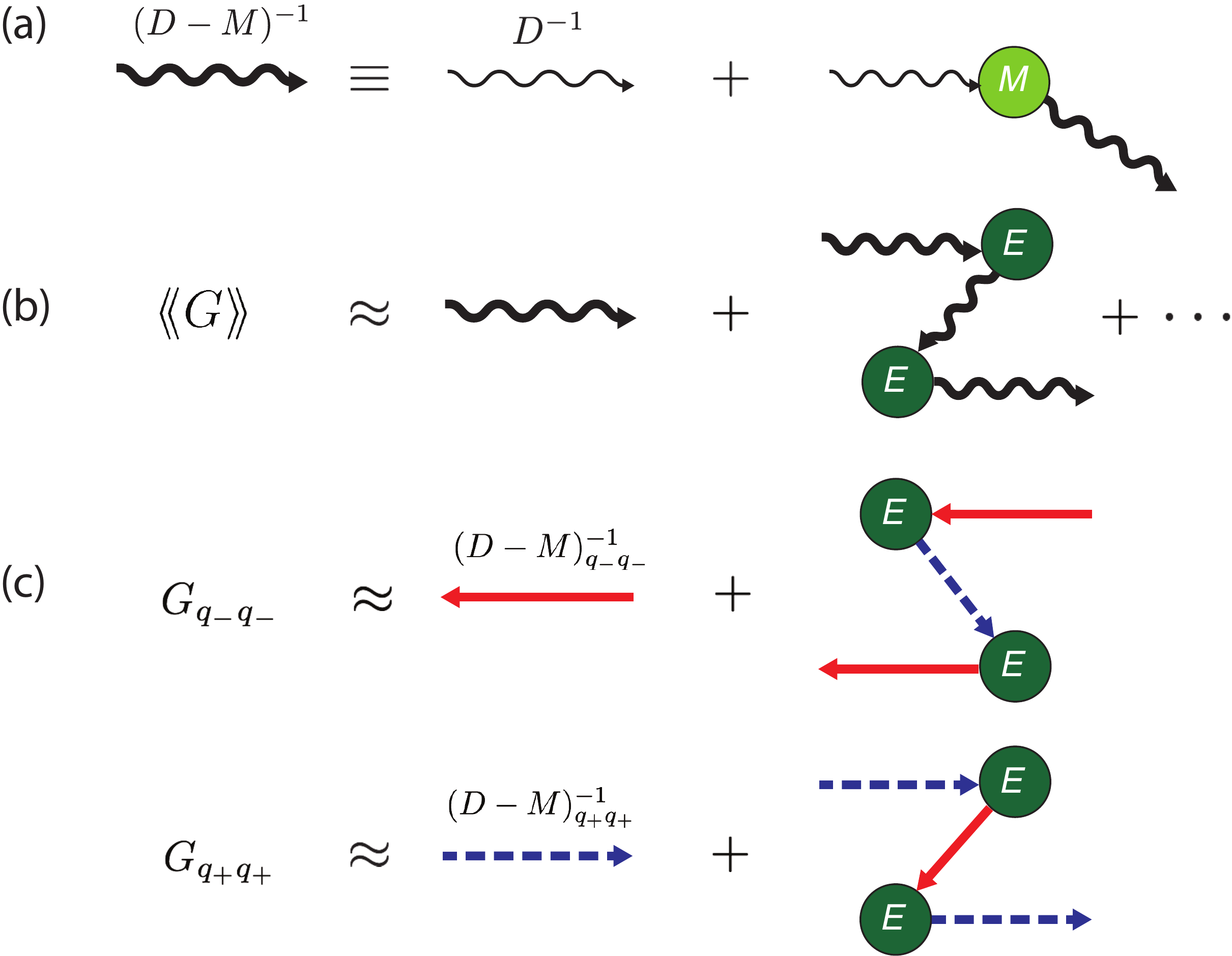}
\caption{Phonon propagator and perturbation expansion. (a) The Dyson equation for the optomechanically modified propagator (thick wavy arrow) due to optomechanical interaction $M$.  (b) The ensemble averaged phonon propagator $\ensemavg{G} $ is the sum of the modified propagator and two successive scatterings $E$ with a modified propagator in between, and higher order contributions. (c) The diagonal element of $\ensemavg{G}$ for backward propagating phonon $q_-$ is the sum of the diagonal element of the modified back propagator (red thick arrow) and the diagonal element of two successive scatterings with a modified forward propagator (blue thick dashed arrow) in between. The result  for forward propagating phonon $q_+$ is also shown below. The asymmetric optomechanical coupling leads to the asymmetry in phonon propagator. }
\label{fig:fig3}
\end{center}
\end{figure}

The random Gaussian noise like scattering potential satisfies 
\be
\ensemavg{E_{q_1 q_2} E_{q_3 q_4}}  = \bar{U}^2 \delta_{q_1+q_3, q_2+q_4}  
\ee
As noted before, when $q= q^{\prime}$, we have $E_{q q^{\prime}} = 0$.  Consequently, we avoid a pole when performing summation or integration. To find the linear response of the phonons, we calculate the diagonal element of the matrix $\tilde{M}$. We focus on the second term in Eq.~ (28), which in more explicit form is 
\bea
 \ensemavg{E \frac{1}{D - M} E}_{ii}   &= &\sum_{j, k}\ensemavg{E_{ij} E_{ki}}\left( \frac{1}{D - M}\right)_{jk}  =
  \bar{U}^2  \sum_{j\neq i } \left( \frac{1}{D - M}\right)_{jj}. 
\eea
The corresponding diagram for calculation is also shown in Fig.~3. 
With this result, it is now much easier to calculate the linear response of phonons, since we get rid of the ensemble average over the random scattering. 
This approach, a weak disorder expansion, neglects multiple scattering events that, in higher dimensions, cause weak localization and weak anti-localization. 
In the high temperature limit that we work with here, we expect thermal fluctuations  and phonon-phonon scattering to lead to relatively fast changes of the effective disorder potential, rendering these multiple scattering events incoherent. At very low temperatures, these effects should be properly included.

\subsection{Analytical results}
To progress further, we can also find the exact form of the $(D-M)^{-1}$ matrix. Specifically,  the matrix $M$ could be rewritten as 
\be
M_{q q^{\prime}}(\omega) = \frac{c_{\mr{cl}}^2 q q^{\prime} f(q) f^{*}(q^{\prime})}{(-\Delta-\omega) - i \kappa/2}  = \frac{\bra{q} \ket{\tilde{\phi}} \bra{\tilde{\phi}}  \ket{q^{\prime}} }{(-\Delta-\omega) - i \kappa/2} 
\ee
where 
\be \ket{\tilde{\phi}} = (\cdots,~c_{\mr{cl}} q f(q),~\cdots)^T,
\ee
is a vector in the space of normal modes that describes the optomechanically modified mode superposition. We  normalize $ \ket{\tilde{\phi}}$ by defining $ \ket{\phi} = \ket{\tilde{\phi}} /N$ with the normalization factor 
\be
N^2 = \left\langle \tilde{\phi} \left\lvert  \right. \tilde{\phi}\right \rangle =  \sum_q c_{\mr{cl}}^2 q^2 |f(q)|^2. 
\ee
We get $M = \lambda P$, with $P = \ket{\phi}\bra{\phi}$ a rank-1 projector and 
\be
\lambda = \frac{N^2}{(-\Delta - \omega)  - i \kappa/2}. 
\ee
This gives the useful identity:
\be
P D^{-1} P = g P
\ee
with $g = \bra{\phi} D^{-1} \ket{\phi}$. Thus the series
\begin{align}
G_M \equiv \frac{1}{D-\lambda P} &= D^{-1} + \lambda D^{-1} P D^{-1} + \lambda^2 D^{-1} P D^{-1} P D^{-1} + \ldots \\
&= D^{-1} + \lambda D^{-1} P D^{-1} + \lambda^2 g D^{-1} P D^{-1} +  \lambda^3 g^2 D^{-1} P D^{-1} + \ldots \\
&= D^{-1} + \lambda \left(\sum_{n=0}^\infty (\lambda g)^n \right) D^{-1} P D^{-1}  \\
&= D^{-1} + \frac{1}{\lambda^{-1} - g} D^{-1} P D^{-1} 
\end{align}

We  now  estimate $g$ in 1-dimension. We have, near the resonance in $f(q)$,
\be
g(\omega) = \frac{\sum_q c_{\mathrm{cl}}^2 q^2 |f(q)|^2 \frac{1}{\Omega_q - \omega - i \gamma_q/2}}{N^2} \approx \frac{(\pi/2) i \rho(\omega) c_{\mathrm{cl}}^2 q_{\omega}^2 |f(q_\omega)|^2 }{N^2}
\ee
where the resonance value $\Omega_{q_\omega}  = \omega$, and phonon density of states $\rho(\Omega)$. We have used the usual delta function description of $1/(x - i \epsilon) = \pi i \delta(x) + \mathcal{P}$ (the Sokhotski-Plemelj theorem), and the factor of $1/2$ comes from the integration of the $\abs{f(q)}^{2}$ function. 
We only keep the contribution from $q$ near the phase-matching condition, preventing a second pole for the non-coupled (back propagating) sound from contributing. We remark that $g$ is purely imaginary, and acts to effectively broaden the value $\kappa$ by an amount $\tilde{\kappa} - \kappa = -2iN^2 g = \pi \rho(\omega) c_{\mathrm{cl}}^2 q_{\omega}^2 |f(q_\omega)|^2$, consistent with our original definition of $f(q)$ including phonon damping Eq.~(13).

Let us consider the prefactor $(1/\lambda - g)^{-1}$ in more detail. Writing it out, we have
\be
\eta \equiv \frac{1}{
1/\lambda - g} \approx \frac{N^2}{(-\Delta - \omega) - i \tilde{\kappa}/2}\ . 
\ee
When the optical power is small, $N^2 \rightarrow 0$, as expected.  As the optical power becomes large, $\tilde{\kappa} \gg \kappa$, and there is an effective increase in the range over which these effects can occur. Note that $N^2/c_{\mathrm{cl}}^{2} |f(q)|^2$ does not change with power, so $\eta \rightarrow i \frac{2N^2}{\pi \rho(\omega) c_{\mathrm{cl}}^2 q_{\omega}^2 |f(q_\omega)|^2}$. In general, the prefactor $\eta$ is in the positive imaginary part of the complex plane.

From this result, we can look again at
\begin{align}
G = \frac{1}{D - \lambda P - E} &= G_M + G_M E G \\ 
&=G_M + G_M E (G_M + G_M E G) 
\end{align}
Putting in the average over disorder, we have
\begin{align}
\ensemavg {G} &= G_M + G_M \ensemavg{E G_M E G} \\
&\approx  G_M + G_M \ensemavg{E G_M E } \ensemavg{G} \\
\Rightarrow \ \ G_M^{-1} \ensemavg{G} &= \mathbb{I} + \ensemavg{E G_M E} \ensemavg{G} \\
\ensemavg{G} &= \frac{1}{G_M^{-1} - \ensemavg{E G_M E}}
\end{align}
The factorization of the mean value in Eq.~(45) is the key point here, and the only approximation.  We note that this approximation is the same one used in the prior resummation in Eq.~(28), but in  more explicit form.

Evaluating this, we define $\ensemavg{G} = (D- \tilde{M})^{-1}$, as before with
\begin{align}
\tilde{M} &= \lambda P + \ensemavg{E D^{-1} E} + \frac{1}{1/\lambda - g} \ensemavg{E D^{-1} P D^{-1} E} \\
&= M + \Sigma_D + \Sigma_P
\end{align}
where $\Sigma_D$ is the self-energy term that leads to diffusion in the absence of the optical field, and $\Sigma_P$ is the new term that will cancel some of this self-energy, leading to reduced diffusion, while $M$ adds (for forward propagating phonons) direct damping via optical cooling. Intuitively, we can understand the $\Sigma_D$ term to correspond to phonons that scatter backwards, then forwards again an indeterminate time later, which leads to a random walk and diffusive transport. On the other hand, the $\Sigma_P$ term has, during the backward propagation stage, a chance to be absorbed (converted into an optical photon and lost). This induced imaginary component in the reflection can, in analogy to the quantum Zeno effect, prevents the backscattering in the first place, as we now show analytically.

We look first at the regular diffusion term $\Sigma_D$. Using the averaging of $\ensemavg{E_{qq'} E_{kk'}} = \bar{U}^2 \delta_{q+k,q'+k'}$, 
\be
\ensemavg{E D^{-1} E}_{qk} = \bar{U}^2 \sum_{k'} (D^{-1})_{q-k+k',k'}
\ee
The diagonal nature of $D$ in the plane-wave basis means that $q - k = 0$ is the only contribution. Thus the effect is purely diagonal, and given by $\bar{U}^2 \sum_{k'} (D^{-1})_{k',k'} \approx 2\pi i \rho(\omega) \bar{U}^2$, where the factor of 2 comes from the degenerate phonons modes propagating in opposite directions. This is the loss via diffusion that we expect from disorder.

Turning now to the optical correction to diffusion $\Sigma_P$, we define
\be
-W_{q-k} \equiv \ensemavg{E D^{-1} P D^{-1} E}_{qk} = \bar{U}^2 \sum_{k'} (D^{-1} P D^{-1})_{q-k+k',k'}
\ee
Looking at the backward propagating modes with $q$ such that $f(q) \approx 0$, we can neglect the $M$ term for the diagonal contributions.  We can also approximate $D^{-1} P D^{-1}$ with terms that correspond only to those near $q_0$, the ones with $f(q_0)$ largest -- everything else is suppressed. That is, we can take $P$ to be a width $w$ approximation to a delta function near $q = q_0$, which means only terms in the sum with $|k' - q_0| < w$ and $|q -k + k' - q_0| < w$ contribute. This leads to an effect for a narrow range of $|q-k| \lesssim w$. Working near resonance, $D^{-1}$ is nearly purely imaginary; the product is thus real and negative. So we expect the overall term to be of order $-\bar{U}^2 W$ with $W \sim |g|^2$. Note that the minus sign, combined with the fact that $\eta$ is the upper half plane, means that this effect strictly decreases the diffusion term near the resonance.

We can bound this by noting:
\be
-W_0 =  \sum_{k'} (D^{-1} P D^{-1})_{k',k'} = \mathrm{Tr}[D^{-1} P D^{-1}]  = \bra{\phi} D^{-2} \ket{\phi} 
\ee
We can evaluate this, in principle to get a (frequency-dependent) contribution. For specificity, let us consider a simplified case where $|q f(q)|^2$ is a lorenztian of width $\gamma/v$ about $q_{c} = \Delta k =  k^{\prime} - k$ and maximum of $q_c^2$.  We also take $\rho = r_{0}/v$. Explicit integration gives 
\begin{align}
N^2 &= c_{\mathrm{cl}}^2 q_{c}^{2} \rho \frac{\pi \gamma}{2} \\
g &= \frac{i\gamma/4}{(\omega - \Omega_c)^{2} + \gamma^{2}/4} \\
W_0 & = \frac{1/4}{(\omega - \Omega_c)^{2} + \gamma^{2}/4}
\end{align}

Looking near resonance, for small $\rho$, we recover for $q \approx - q_c$, and $\Delta = -\Omega_c$, 
\bea                                                                                                                                                                                                                                                                                                                                                                  
\frac{\tilde{M}_{qq}}{D_{0}q^2} &\approx & i \left[ 1 - \frac{1}{2\pi \rho \bar{U}^{2} }\frac{2N^{2}}{\kappa -2i gN^{2}   } \bar{U}^{2 }W_0 \right] \\
&=& i\left[1 - \frac{c_{\mathrm{cl}}^{2} q_{c}^{2} \gamma/2 }{\kappa + c_{\mathrm{cl}}^{2}  q_{c}^{2}\pi \rho}  \frac{1}{\Gamma^{2}} \right]  \\
&=&  i\left[1 - \frac{c_{\mathrm{cl}}^{2} q_{c}^{2}/(2\gamma\kappa)  }{1 + c_{\mathrm{cl}}^{2}  q_{c}^{2}/(\gamma\kappa) \cdot \pi \rho \gamma}  \right] 
\eea
normalized with the laser-free diffusion constant $D_{0} = 2\pi \rho \bar{U}^2 / q^2$. At high power (large $c_{\mathrm{cl}}^{2}$), this further simplifies to 
\be
D/D_{0} \approx 1 - \frac{1}{2\pi \rho \gamma}
\ee
That is, the diffusion is reduced, but the reduction is limited by the number of states that fit in the width of the phase matching, $\sim \rho\gamma$. If only a few states phase match, the reduction in diffusion could become large.

\subsection{Comparison between analytical formula and numerical results}
We can compare the approximate formula of Eq.~(57) for the center mode of the back-propagating (CCW) phonon bands,  with the numerical results based on Eq.~(29), as shown in Fig.~4. The forward-propagating (CW) phonons are more strongly coupled to the optical mode, so their linewidth is dominated by optomechanical cooling, as indicated by the large value of the mode-matching function $\abs{f(q)}^2$. 

The numerical data points (in circles) are calculated using Eq.~(29) and then fitted by curves of the form $y = 1 - a\cdot x/(1+ b\cdot x)$, which shows the dependence of the phonon diffusion on rescaled pump power. From our formula Eq.~(57), we expect $a \to 1/2$ and $b/\pi \to \rho \gamma$.  The lease square best fit values for $a, b$ are consistent with our theory prediction. Since our formula is obtained by making approximations using the continuum theory, there is some minor discrepancy between the fitted values of $a$ and the coefficients in the formula. 

We see that as the pump power increases, the diffusion of the CCW phonon keeps decreasing but saturates eventually.  The damping of back-propagating (CCW) phonon mainly comes from its scattering with disorder/impurity of the material, and it is also not directly modified by the optics, since the mode-matching function $\abs{f(q)}^2$ is near zero, as shown below. The decrease in diffusion comes from a decrease in the second term of Fig.~3(c), where two scatterings happen with a modified forward phonon propagator in between.  Intuitively, the random walk nature of diffusion is modified by the back scattering being suppressed -- it prevents those terms from contributing to the phonon motion. This is a classical analog of Zeno's paradox.

\begin{figure}[ht!]
\begin{center}
\includegraphics[width=0.7\textwidth]{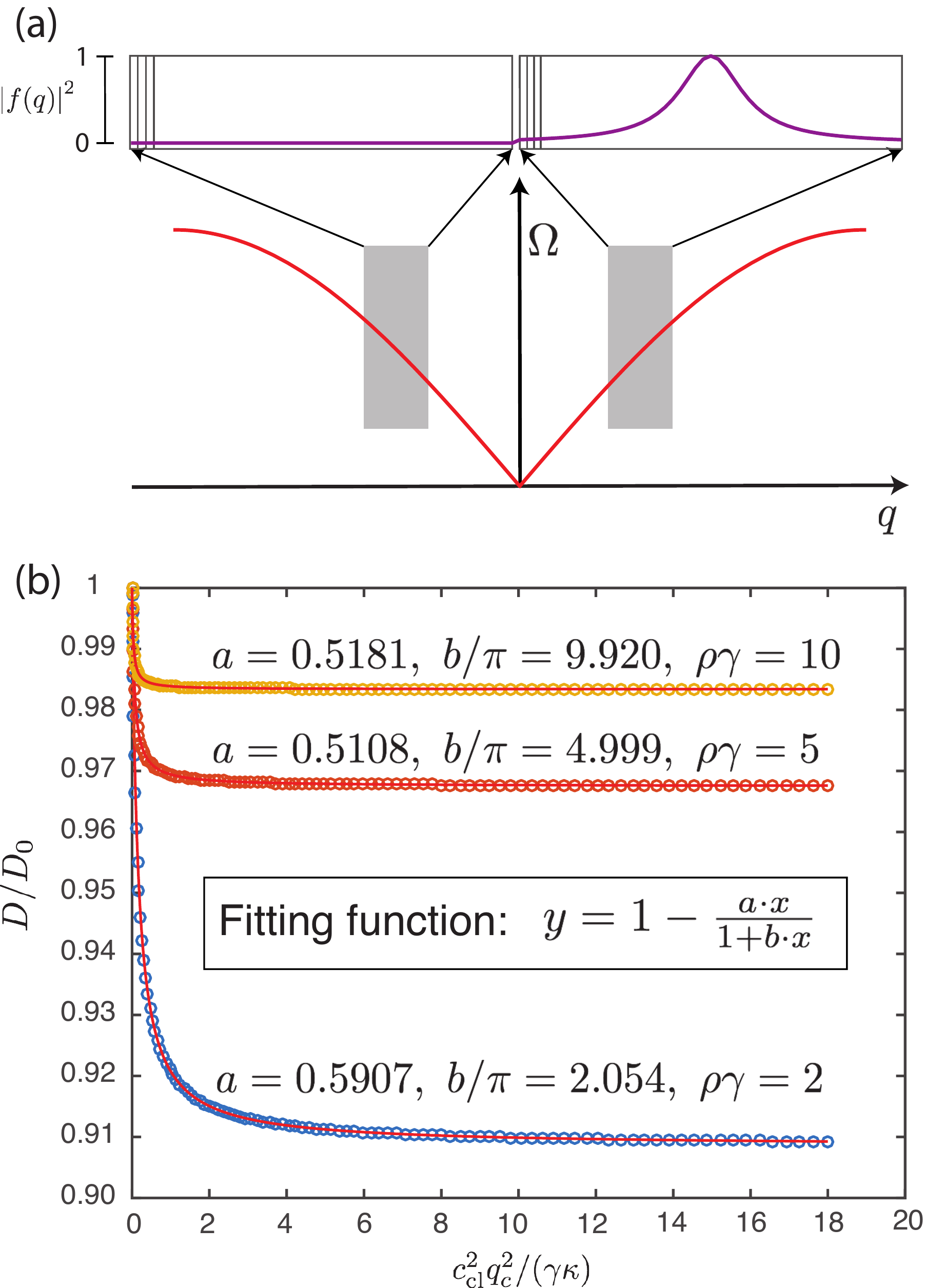}
\caption{(color online). (a) We symmetrically choose equally spaced phonon modes for both $q>0$ and $q<0$ in the long wavelength linear dispersion regime. The phase-matching function $\abs{f(q)}^2$ shows the highly asymmetric coupling to optics between CW and CCW phonons.  (b) We choose phonon mode spacing $\Delta q = 0.0001$, such that density of states $\rho = 10000$. We then choose mechanical damping $\gamma = 0.0002, 0.0005, 0.001$, which gives $\rho\gamma = 2, 5, 10$, for blue, red and orange data, respectively. We plot the normalized diffusion as a function of normalized pump power, and then fit the data points with the function in the inset. The resulting analytical comparison shows close agreement. }
\label{fig:fig4}
\end{center}
\end{figure}

\section{Conclusion}
We provide a general model of phonon propagation with the presence of directional optomechanical interaction and random phonon-phonon scattering via impurities. We perform both numerical simulation of the optomechanically-modified phonon propagation in the presence of disorder and perturbative calculation of phonon propagator to find the linear response of the phonons. Our result shows an increase of the linewidth of phonons directly coupled to the optics and a decrease of the linewidth of phonons propagating in the opposite direction as optical pump power increases. Furthermore, we see a direction-dependent decrease of the diffusion constant of the material, due to back-scattering to optomechanically modified phonons. These results suggest that in Brillouin scattering, high $Q$ optical experiments, the laser pump may induce chiral behavior for the mechanical systems even in the continuum limit, at least over the phase-matching bandwidth.

\section{Acknowledgements} 
We thank G. Bahl, S. Kim, P. Hao and S. Papp for helpful discussions. This work was performed in part during a stay at the NSF funded KITP in Santa Barbara (under grant No. NSF PHY-1125915) and by the Physics Frontier Center at the JQI.

\bibliography{ChiralPhonon_v16}
\end{document}